# Hybrid amplitude determination in charged kaon photoproduction and comparison with partial waves analysis in the energy range of 1800 to 2300 MeV

Gh.Forozani

Department of Physics, Bu-Ali Sina University, Hamadan,I. R. Iran
Email: forozani@basu.ac.ir

Abstract

The magnitude of four independent amplitudes are determined in charged kaon photoproduction in the energy range of 1500 to 2300MeV incident photon. Differential cross section and three polarization parameters are required for such amplitudes reconstruction at different kaon scattering angles. This analysis has indicated that the magnitudes of spin-flip amplitudes of the target nucleon dominate over the non-flip ones in all energies and scattering angles. The results have been compared with the solution of phase shift analysis sp98 at all energies. This analysis indicates that here is a good agreement between the present work and the results of phase shift analysis at all energies and scattering angles.

## 1. Introduction

The use of polarization experiments as a source of information about particle reactions in atomic, nuclear and high energy physics have increased substantially in recent years[1-8]. For the case of kaon photoproductiona certain ensemble of observables must be measured. If these observables contains sufficient information for an exhaustive description of the interaction it is called complete[9-10]. From which a direct reconstruction of the scattering matrix element can be performed in a unique way, without any theoretical input, except the symmetry properties. The non-dynamical spin polarization analysis of the complex reaction amplitudes not only enables us to obtain dynamical information, it is also capable to checking the validity of conservation laws such as time reversal, parity and identical particles in hadronic interactions[11]. The best way for checking the validity of model-dependent contributions to phase shift analysis is to compare the phase shift analysis amplitude predictions with the direct reconstruction from the amplitude analysis. From this point of view, the amplitude analysis is to be considered as a complementary method with respect to the phase shift analysis. The results agreement will support the phase shift analysis which may be used to predict conveniently unmeasured quantities. The consistency between the results will either suggest a possible anomaly in the data base, or will throw a doubt on the validity of the obtained phase shift analysis solutions[12].

In order to obtain the spin amplitudes of an elastic interaction, one has to find their relations to the experimental observables by constructing the interaction matrix. Such a relationship is normally complicated. The way to get rid of such complications is to use the optimal formalism[13-14]. In this formalism the interaction matrix is diagonalized as much as possible without loss of generality. The recent experiments that have been performed in the case of kaon photoproduction, are all incomplete, and this is why the phases of amplitudes are not determined. In this article we have tried to reconstruct the magnitudes of the four independent hybrid reaction amplitudes for the $(1 + 1/2 \rightarrow 0+1/2)$ interaction.

The magnitudes of these amplitudes are obtained in terms of four experimental observables, namely $d\sigma/d\Omega$ (the unpolarized differential cross-section), $P(\theta)$ (the recoil nucleon polarization), $T(\theta)$ (the conventional polarized target asymmetry) and $\Sigma(\theta)$ (the conventional polarized photon asymmetry).

## 2. General Formalism

Lets consider the general interaction

$$S_A + S_B \rightarrow S_C + S_D \,, \quad (1)$$

where $S_A$, $S_B$, $S_C$ and $S_D$ are spin of A, B, C and D particles respectively. In order to obtain the reaction matrix for (1), the standard factorization procedure is used[15] to decompose this reaction into two constituent reactions each with one zero spin particle

$$S_A + 0 \rightarrow S_C + 0$$
$$0 + S_B \rightarrow 0 + S_D, \qquad (2)$$

The compact form of each constituent reaction matrix is given by[14]

$$M_1 = \sum_{l,\lambda} D(l,\lambda) S^{l\lambda}$$

Where $D(l,\lambda)'s$ are spin amplitudes containing all the dynamical information concerning the interaction mechanism and $S^{l\lambda}$ are the spin-momentum tensors corresponding to the particles with spin $S_A$ and $S_C$. In (3), l and $\lambda$ are spin components along quantization axis for each particle. Considering a similar reaction matrix $M_2$ for the second constituent reaction in (2), the general reaction matrix is given as $M_1$.

The constituent reaction matrices for four spin 1/2 particles are given as

$$M_1 = a_0 I + a_1 \sigma_1.l + a_2 \sigma_2.m + a_3 \sigma_3.n \quad (4)$$
$$M_2 = b_0 I + b_1 \sigma_1.l + b_2 \sigma_2.m + b_3 \sigma_3.n, \quad (5)$$

where $\sigma_1$ and $\sigma_2$ are the Pauli matrices and I is the identity matrix. Unitvectors l, m and n are defined as

$$l = (\vec{q}_C - \vec{q}_A)/|(\vec{q}_C - \vec{q}_A)|$$
$$m = (\vec{q}_A \times \vec{q}_C)/|(\vec{q}_A \times \vec{q}_C)|$$
$$n = l \times m, \quad (6)$$

where $q_i s$ are the center of mass momenta for the general reaction (1). The general reaction matrix is given by:

$$M = D_{00} + D_{10}(\sigma_1.l) + D_{20}(\sigma_1.m) + D_{30}(\sigma_1.n) + D_{01}(\sigma_2.l) + D_{02}(\sigma_2.m) + D_{03}(\sigma_2.n) + D_{11}(\sigma_1.l)(\sigma_2.l) + D_{12}(\sigma_1.l)(\sigma_2.m) + D_{13}(\sigma_1.l)(\sigma_2.n) + D_{21}(\sigma_1.m)(\sigma_2.l) + D_{22}(\sigma_1.m)(\sigma_2.m) + D_{23}(\sigma_1.m)(\sigma_2.n) + D_{31}(\sigma_1.n)(\sigma_2.l) + D_{32}(\sigma_1.n)(\sigma_2.m) + D_{33}(\sigma_1.n)(\sigma_2.n). \qquad (7)$$

Each of the constituent reaction in (2) is characterized by an initial density matrix $\rho_i$ and a final density matrix $\rho_f$. Both $\rho_i$ and $\rho_f$ describe the spin polarization states of particles before and after interaction. The overall initial density matrix is given in terms of outer products of the constituent initial density matrices, namely

$$\rho_i = \rho_i^{uv} \otimes \rho_i^{UV}, \qquad (8)$$

and the final density matrix is given in terms of the initial density matrix

$$\rho_f = M(\rho_i^{uv} \otimes \rho_i^{UV})M^\dagger \qquad (9)$$

The experimental observables are given in terms of the expectation values of a certain spin-momentum tensor Q in the final states. The corresponding observable in the final state is given by

$$Q = Q^{\psi\omega} \otimes Q^{\Psi\Omega}, \qquad (10)$$

Where $Q^{\psi\omega}$ is a $(2S_C + 1) \times (2S_C + 1)$ matrix which describes the final state of $S_C$

and $Q^{\Psi\Omega}$ is a $(2S_D + 1) \times (2S_D + 1)$ matrix which describes the final state of $S_D$.

### 3. Optimal conventions

It is time to impose optimal conditions to simplify the relations between the observables and the bilinear combination of the amplitudes. In this formalism the initial and final density matrices as well as the spin momentum tensors are defined in a way such that the elements of the general interaction matrix are as diagonal as possible. The best way is to choose the density matrices and spin momentum tensors to be all hermitian [13-14].

In the case of kaon photoproduction, in addition to Lorentz invariance, the conservation of parity, reduces the number of independent amplitudes to six. For more simplicity we shall use thehybrid frame [16]. In this frame the polarization direction of the photon is in the reaction plane and in the direction of its momentum, but the polarization of nucleon is perpendicular to the reaction plane. Since the photon has only two directions of polarization the number of independent amplitudes is reduced from six to four and only the following spin amplitudes remain: D(+,11) and D(+, 22) are spin non-flip amplitudes whereas D(+, 12) and D(+, 21) are spinflip amplitudes. The numbers 1 and 2 correspond to the positive and negative components of the nucleon spin respectively.

**4. Construction of Transition amplitudes**

Let us define the amplitudes in terms of the two parameters $h_i$ and $\alpha_i$

D(+, 11) = $h_1 \exp(i\alpha_1)$

D(+, 21) = $h_2 \exp(i\alpha_2)$

D(+, 12) = $h_3 \exp(i\alpha_3)$

D(+, 22) = $h_4 \exp(i\alpha_4)$.                                    (14)

The magnitudes of the four independent amplitudes in terms of the experimental observables $\Sigma(\theta), T(\theta), P(\theta)$ and the differential cross section are given as follows:

$|h_1| = \frac{\sqrt{d\sigma/d\Omega}}{2}[1+\Sigma(\theta) + T(\theta) + P(\theta)]^{1/2}$                (15)

$|h_2| = \frac{\sqrt{d\sigma/d\Omega}}{2}[1+\Sigma(\theta) - T(\theta) - P(\theta)]^{1/2}$                (16)

$|h_3| = \frac{\sqrt{d\sigma/d\Omega}}{2}[1-\Sigma(\theta) + T(\theta) - P(\theta)]^{1/2}$                (17)

$|h_4| = \frac{\sqrt{d\sigma/d\Omega}}{2}[1+\Sigma(\theta) + T(\theta) + P(\theta)]^{1/2}$                (18)

The values of these experimental observables are collected from BONN-ELSA and LEPS[17-18]. The magnitudes of four independent amplitudes are unambiguously constructed for the energies 1500, 1800 and 2300 MeV and are plotted versus the kaon scattering angles from 15. to 50. in figures1to3. In these figures we have also compared the results of the present work(p.w.) with the theoretical predictions using the CNS Data Analysis Center sp98 program.[19].

**5. Conclusion**

In addition to symmetry constraints, optimal conventions have reduced the number of independent amplitudes to four, D(+, 12), D(+, 21) and D(+, 11), D(+, 22) which are corresponded to spin flip and spin non-flip amplitudes respectively. Polynomial fitting has been used in all curves. This analysis have indicated that at all measured scattering angles and energy range, the spin flip up-down and down-up dominates over the spin non-flip amplitudes. The magnitudes of amplitudes in the energy range of 1500 to 2300MeV are smooth. In general we can say that the initial orientation of spin polarization of the target nuclei plays a significant role in the probability of spin to flip. We have found that the results of the present work and the results of phase shift analysis sp98 are in good agreements at all energies and scattering angles.

**References**

[1] G.Forozani: ActaPhys. Slov. 54 (2004)1

[2] N. GhahramanyandG.Forozani:Phys.Rev.C 61 (2000) 1086

[3] J. Bystricky,C. Lechanoine-LelucandF. Lehar:Eur.Phys.J.C 4 (1998) 607

[4]C. Lechanoine-LelucandF. Lehar:Rev.Mod.Phys. 65 (1993) 47

[5] M.W. Mcnaughton et al.: Phys.Rev.C 41 (1990) 2809

[6] R. Hausammann et al.: Phys.Rev.D 40 (1989) 22

[7]P. Bydzovsky,T. Mart,Phys.Rev.C76, 065202 (2007)

[8] A. salam,T. Mart, K.Miyagawa.: arXiv:0906.0316 (2009)

[9] L. D. Puzikov, R. M. Rydin and YA. A. Smorodinskii: Nucl. Phys. 3 (1957)436

[10] G. Martin et al.: Nucl.Phys.A 487, (1988) 563

[11] M. Moravcsik:Phys.Rev.D 27 (1983) 289

[12] J. Ball et al.: IL Nuovo Cimento 111 A (1998) 13

[13] G. Goldstein andM. Moravcsik: Ann.Phys.(N.Y.) 98 (1976) 128 4

[14] G. Goldstein and M. Moravcsik: Nuovo Cim.A,84(1984)39

[15]P. Chonka and M.Moravcsik:Phys.Rev.D 1 (1970) 1821

[16] G. Goldstein and M. Moravcsik: Ann. Phys. 142 (1982) 219

[17] R.G.T.Zegers et al.: Phys. Rev. Lett. 91, 092001 (2003)

[18] K. -H. Glander et al.: Eur. Phys. J. A19(2004)251

[19] T. Mart et al.: CNS Data Analysis Center, sp98, http://lux2.phys.va.gwu.edu/analysis/prk analysis.html.

E = 1500 MeV

| $\theta_{cm}$ | D(+, 11) | dD(+, 11) | D(+, 21) | dD(+, 21) | D(+, 12) | dD(+, 12) | D(+, 22) | dD(+, 22) |
|---|---|---|---|---|---|---|---|---|
| 15 | 0.126 | 0.014 | 0.343 | 0.040 | 0.240 | 0.058 | 0.202 | 0.012 |
| 20 | 0.097 | 0.015 | 0.356 | 0.030 | 0.239 | 0.056 | 0.187 | 0.019 |
| 25 | 0.071 | 0.018 | 0.363 | 0.042 | 0.239 | 0.061 | 0.177 | 0.018 |
| 30 | 0.062 | 0.023 | 0.372 | 0.043 | 0.241 | 0.062 | 0.167 | 0.019 |
| 35 | 0.059 | 0.027 | 0.392 | 0.039 | 0.231 | 0.056 | 0.164 | 0.016 |
| 40 | 0.074 | 0.027 | 0.381 | 0.038 | 0.225 | 0.051 | 0.159 | 0.012 |
| 45 | 0.085 | 0.035 | 0.378 | 0.044 | 0.217 | 0.049 | 0.151 | 0.013 |
| 50 | 0.099 | 0.031 | 0.372 | 0.045 | 0.208 | 0.047 | 0.146 | 0.008 |

Table 1: The magnitudes of the four independent amplitudes in the hybride frame for different values of kaon center of mass scattering angles and incident photon energies for charged kaon photoproduction.

E = 1800 MeV

| $\theta_{cm}$ | D(+, 11) | dD(+, 11) | D(+, 21) | dD(+, 21) | D(+, 12) | dD(+, 12) | D(+, 22) | dD(+, 22) |
|---|---|---|---|---|---|---|---|---|
| 15 | 0.050 | 0.010 | 0.318 | 0.040 | 0.224 | 0.021 | 0.157 | 0.007 |
| 20 | 0.070 | 0.019 | 0.339 | 0.042 | 0.231 | 0.022 | 0.150 | .008 |
| 25 | 0.105 | 0.035 | 0.351 | 0.045 | 0.241 | 0.027 | 0.151 | 0.009 |
| 30 | 0.120 | 0.023 | 0.358 | 0.038 | 0.248 | 0.019 | 0.155 | 0.011 |
| 35 | 0.140 | 0.022 | 0.366 | 0.034 | 0.249 | 0.026 | 0.166 | 0.008 |
| 40 | 0.130 | 0.027 | 0.367 | 0.041 | 0.234 | 0.023 | 0.177 | 0.008 |
| 45 | 0.130 | 0.027 | 0.361 | 0.042 | 0.238 | 0.024 | 0.174 | 0.009 |
| 50 | 0.110 | 0.029 | 0.341 | 0.037 | 0.238 | 0.019 | 0.187 | 0.011 |

Table 2: The magnitudes of the four independent amplitudes in the hybride frame for different values of kaon center of mass scattering angles and incident photon energies for charged kaon photoproduction.

E=2300 MeV

| $\theta_{cm}$ | D(+, 11) | dD(+, 11) | D(+, 21) | dD(+, 21) | D(+, 12) | dD(+, 12) | D(+, 22) | dD(+, 22) |
|---|---|---|---|---|---|---|---|---|
| 15 | 0.117 | 0.026 | 0.362 | 0.009 | 0.224 | 0.010 | 0.091 | 0.023 |
| 20 | 0.174 | 0.020 | 0.395 | 0.009 | 0.242 | 0,011 | 0.098 | 0.029 |
| 25 | 0.205 | 0.019 | 0.420 | 0.008 | -.255 | 0.012 | 0.109 | 0.033 |
| 30 | 0.209 | 0.019 | 0.424 | 0.008 | 0.261 | 0.013 | 0.117 | 0.035 |
| 35 | 0.195 | 0.012 | 0.403 | 0.006 | 0.261 | 0.009 | 0.119 | 0.024 |
| 40 | 0.167 | 0.012 | 0.358 | 0.005 | 0.256 | 0.009 | 0.112 | 0.023 |
| 45 | 0.137 | 0.011 | 0.301 | 0.005 | 0.246 | 0.009 | 0.096 | 0.023 |
| 50 | 0.127 | 0.010 | 0.276 | 0.004 | 0.234 | 0.008 | 0.090 | 0.022 |

Table 3: The magnitudes of the four independent amplitudes in the hybride frame for different values of kaon center of mass scattering angles and incident photon energies for charged kaon photoproduction.

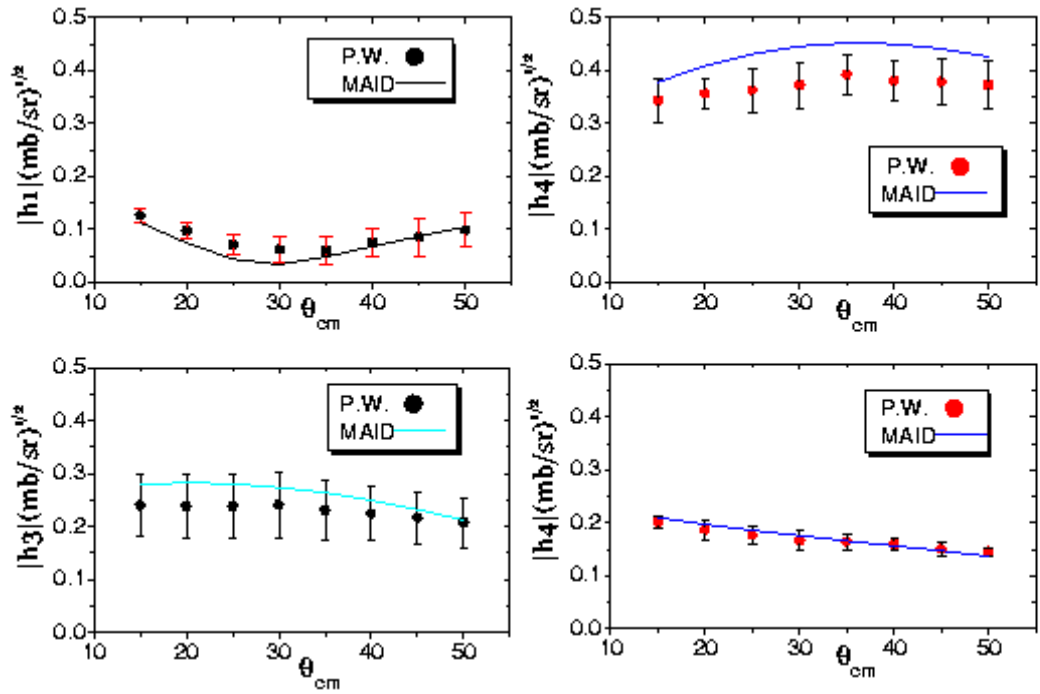

Figure 1: The magnitudes of four independent amplitudes as a function of kaon center of mass scattering angles at fixed photon energies (1500 MeV) for charged kaon photoproduction.

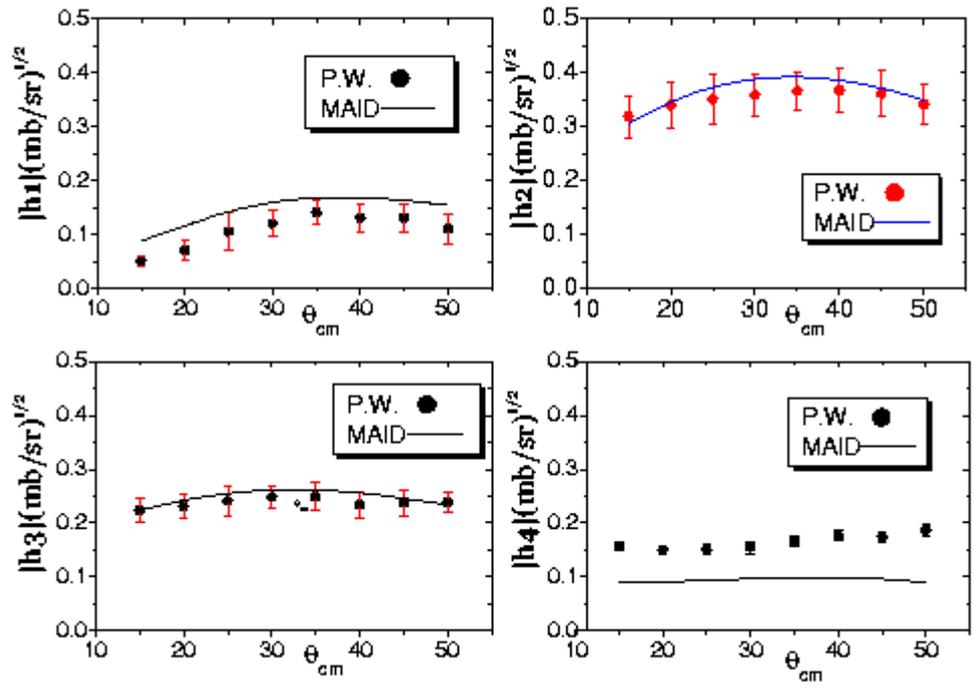

Figure 2: The magnitudes of four independent amplitudes as a function of kaon center of mass scattering angles at fixed photon energies (1800 MeV) for charged kaon photoproduction.

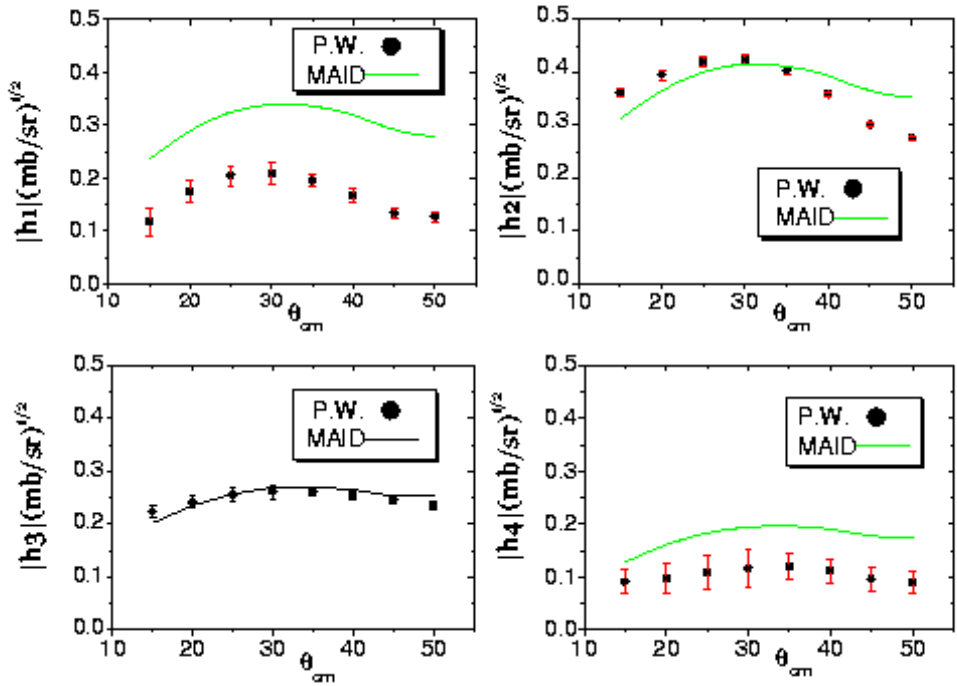

Figure 3: The magnitudes of four independent amplitudes as a function of kaon center of mass scattering angles at fixed photon energies (2300 MeV) for charged kaon photoproduction.